# Wetting of soap bubbles on hydrophilic, hydrophobic and superhydrophobic surfaces

Steve Arscott

*Institut d'Electronique, de Microélectronique et de Nanotechnologie (IEMN), CNRS UMR8520,*

*The University of Lille, Cité Scientifique, Avenue Poincaré, 59652 Villeneuve d'Ascq, France*

Wetting of sessile bubbles on solid and liquid surfaces has been studied. A model is presented for the contact angle of a sessile bubble based on a modified Young's equation - the experimental results agree with the model. A hydrophilic surface results in a bubble contact angle of 90° whereas on a superhydrophobic surface one observes 134°. For hydrophilic surfaces, the bubble angle diminishes with bubble radius - whereas on a superhydrophobic surface, the bubble angle increases. The size of the Plateau borders governs the bubble contact angle - depending on the wetting of the surface.

An understanding of the behaviour of bubbles, liquid films, foams and froths is vital for several fields including mining, manufacturing, materials, security and food production.[1] In addition, the use of soap bubbles and films has been recently demonstrated in micro[2,3] and nanotechnologies.[4,5] Soap bubbles – surely one of nature's most beautiful objects – and films have been studied for some time now;[6-9] more recent investigations include *inter alia* their composition,[10] organisation,[11,12] electrification,[13-15] magnetization,[16] wetting,[17-19] stability[20] and mechanical[21] and optical properties.[22] Here, the wetting of sessile soap bubbles resting on solid surfaces (hydrophilic to superhydrophobic) and on a liquid surface is studied. The results have potential implications in the aforementioned applications.

Fig. 1 shows a sessile droplet and a sessile bubble resting on a solid surface. For the droplet [Fig. 1(a)], the balance between the surface tension of the liquid $\gamma_l$, the solid-liquid $\gamma_{sl}$ and the solid-vapour $\gamma_{sv}$ surface energies - at the liquid/solid/vapour interface - is given by Young's equation[23,24] and leads to a liquid contact angle $\theta_l$:

$$\gamma_{sv} = \gamma_l \cos\theta_l + \gamma_{sl} \qquad (1)$$

For a sessile bubble [Fig. 1(b)], Young's equation needs to be modified to take into account the internal surface of the bubble. As the bubble has two surfaces, its effective surface tension $\gamma_b$ is twice the surface tension of the liquid, i.e. $\gamma_b = 2\gamma_l$. In addition to this, an extra $\gamma_l$ term is required - acting in the same direction as $\gamma_{sl}$ – in order to take into account creation of bubble surface inside the bubble. Thus, the modified Young's equation for an ideal sessile bubble forming a contact angle $\theta_{b0}$ with a solid surface can be written as:

$$\gamma_{sv} = 2\gamma_l \cos\theta_{b0} + \gamma_l + \gamma_{sl} \qquad (2)$$

Eq. (1) and Eq. (2) allow us to write the contact angle of an ideal sessile bubble $\theta_{b0}$ resting on a surface in terms of the contact angle of a droplet of bubble solution $\theta_l$ resting on the same surface:

$$\cos\theta_{b0} = \frac{1}{2}(\cos\theta_l - 1) \qquad (3)$$

Eq. (3) assumes that the liquid film thickness at the bubble-solid interface is of the order of the liquid film forming the bubble and that the droplet and bubble are considered to be large enough so that the diminishing contact angle with droplet radius effect[25] – controversially attributed to the line tension[24] – is negligible.

A commercially available soap solution (Pustefix, Germany) was used to generate bubbles for the experiments - a soap solution is a mixture of pure water, a thickener (e.g. glycerol) and a surfactant (e.g. an organosulphate). The surface tension of the solution was measured to be 28.2 mJ m$^{-2}$ (standard deviation = 0.3 mJ m$^{-2}$) using the pendant drop method[26] and applying the appropriate correction factor[27] – a value comparable with other experiments concerning soap bubbles and films.[13,17,18] As a calibration measurement, deionized water was measured – the result was a surface tension of 72.7 mJ m$^{-2}$ (1.2). The density of the solution was measured to be ~ 1000 kg m$^{-3}$.

The different wetting surfaces were fabricated using polished silicon wafers (Siltronix, France). In order of decreasing wetting – 'Surface A' is a 200 nm thick layer of silicon dioxide grown on a silicon wafer using wet thermal oxidation. 'Surface B' is a ~ 100 µm thick PDMS layer – Sylgard® 184 (Dow Corning, USA) spin-coated onto a silicon wafer. 'Surface C' is a ~250 nm Teflon layer obtained using spin-coating of Teflon® AF 1600 (Dupont, USA) diluted with Fluorinert FC-75 (3M, USA).[28] 'Surface D' is composed of "black silicon"[29] produced using dry etching – this surface was subsequently deposited with a ~20 nm thick fluorocarbon layer using a $C_4F_8$ plasma (STS, UK).

The measured contact angles $\theta_l$ of the soap solution on the surfaces A-D are given in the Table. As the values of $\gamma_{sv}$ are very well known for Teflon[24] (15±2 mJ m$^{-2}$) and PDMS[30] (23.5±1.5 mJ m$^{-2}$) we can determine the values of $\gamma_{sl}$ to be (1.8±2.5 mJ m$^{-2}$) and (6.5±2 mJ m$^{-2}$). For a superhydrophobic surface, $\gamma_{sv}$ is of the order of 5-10 mJ m$^{-2}$ which indicates that $\gamma_{sl}$ is in the range 15.5-20.5 mJ m$^{-2}$. For the wetting measurements on a liquid film, Surface A was dipped into the solution to form a uniform liquid film prior to deposition of the bubble from the pipette. Bubbles were generated for the experiments using a pipette (Bio-Rad, France) having a tip diameter of ~0.5 mm. This pipette produced bubbles having radii in the 0.5-10 mm range. All surface preparation and experiments were performed in a class ISO 5/7 cleanroom ($T$ = 20°C±0.5°C; $RH$ = 45%±2%). The data was gathered using a commercial Contact Angle Meter (GBX Scientific Instruments, France).

Bubbles of were deposited onto surfaces A-D in order to form sessile bubbles [Fig. 2]. Millimetre-sized soap bubbles are quasi-spherical with radius of curvature $R$ [Fig. 2(a)] and a contact angle $\theta_b$ [Fig. 2(b)] as the Bond number[24] is small. When depositing a bubble onto a solid surface, despite the presence of the thickener, their drainage can cause the formation of liquid layer $h$ (the Plateau border)[9] [Fig 2(c)] at the bubble-solid interface–larger than the soap film thickness even approaching the bubble base radius $r$ [Fig 2(d)].

Fig. 3(a) shows plots the bubble contact angle $\theta_b$ versus $r$ in the range 1.5 to 3.5 µm.[18] For the hydrophilic and hydrophobic surfaces, the value of $\theta_b$ reduces with reducing $r$ whereas for the superhydrophobic surface the value of $\theta_b$ increases with reducing $r$. This latter observation appears to call into question a "line tension" explanation[17] for the effect – as the line tension should always act to reduce the contact angle for diminishing $r$.

Fig. 3(b) plots $\theta_b$ versus a dimensionless ratio of two lengths $h/R$ associated with the bubble.[24] The standard deviation of the data points was determined to be 1.8°. For a given surface, the measured value of $\theta_b$ – corresponding to small values of $h/R$ – increases from surfaces A to D, i.e. decreased wetting; as is the case with droplets. As $h/R$ increases, the value of $\theta_b$ decreases - the data suggesting a near-linear relationship[19] for the four solid surfaces tested in the range $h/R$ = 0 to 0.7. A linear fit allows us to calculate the intercept bubble angle $\theta_{bi}$ using extrapolation. The value of $\theta_{bi}$ can be compared with the theoretical value of $\theta_{b0}$ predicted by Eq. (3) as $h/R \to 0$ – see the Table - within measurement error, the values compare rather well for all surfaces tested. A large variation of the bubble contact angle is observed here for the soap bubbles on superhydrophobic surfaces (134° → 76° - Fig. 3(b) open squares) and on hydrophilic surfaces (88.3° → 45.4° - Fig. 3(b) open triangles) – the observations can be compared to those of Rodrigues et al[17] who observed relatively small variations of the contact angle of a sessile soap bubble resting on a wet (~4°) and dry (<10°) surfaces and who invoked the controversial[24] line tension effect[17] to explain their observations.

Eq. (3) can explain the measured value of $\theta_b \to \theta_{b0}$ as $h/R \to 0$ i.e. the wetting contact angle of an ideal bubble on a solid surface – which, as we have seen, is not equal to 90° for surfaces where $\theta_l$ > 0. However, in order to explain why $\theta_b$ decreases when $h/R$ is increased we can implement a first-order solution of an analytical model[19] for sessile bubbles wetting a solid surface. By using Eq. (3) – where $h/R \to 0$ – together with the analytical solution given in Ref. [19], we can compute the values of $h/R$ which correspond to a value of $\theta_b$; these are shown as dashed lines in Fig. 3(b). The solutions correspond well with the experimentally obtained values. The measured slopes, $\alpha = d\theta_b/d(h/R)$, correspond well to those predicted by the model (the near-linear range was taken to be $h/R$ = 0 → 0.4). The model[19] predicts a minimum slope $\alpha$ at $\theta_l$ = 90° - this is consistent with the experimental data, see the Table. However, it should be noted that as $h/R$ approaches unity for bubble sizes studied here – gravity (the capillary length of the soap solution is 5.36 mm) is likely to deform the Plateau borders resulting in a reduced $h$ and a larger $R$ – this provides an explanation for the data points as $h/R \to 1$.

In terms of the bubble film thickness, no measurement of this was performed but colors[6-8] are visible directly after bubble deposition indicating an initial film thickness in the sub-micrometre range[24] – drainage and evaporation produces black regions[6] (thickness < 10nm)[24] where presumable the bubble first bursts[20] after a lifetime of seconds to tens of seconds. In addition, it is not the objective of the current article, but there is also the question as to whether or not a continuous liquid film is present at the bubble-solid interface – irrespective of the value of $h/R$. For an ideal bubble wetting a solid surface, the schematic diagram in Fig. 1(b) indicates an ideal bubble and that the film at the bubble-surface interface is continuous. However, Eq. (2) and Eq. (3) simply imply that a single bubble surface needs to be created inside the bubble, i.e. the extra $\gamma_l$ term on the r.h.s. of Eq. (3). Experimentally, as the sessile bubbles have a finite lifetime (~seconds to tens of seconds), bubble bursting[20] and its outcome can be observed. It was observed that one of two outcomes can be the result from bubble bursting – (i) the bubble bursts leaving a flat film of liquid (of radius $r$) which contracts into a well-defined spherical droplet having a contact angle $\theta_l$ and (ii) the bubble bursts resulting in a liquid ring of radius $r$.[20] This ring is observed to be unstable and either coagulates into a single droplet – as with the latter case (i), indicating a continuous film - or breaks-up into smaller droplets having a contact angle $\theta_l$ – presumably due to a Rayleigh-Plateau instability - Fig. 4 shows these outcomes for small values of $h/R$. In general for a large $h/R$ ratio (i) is always observed, however as $h/R$ reduces then, in general, bursting on a superhydrophobic and hydrophobic surface [Figs. 4(a) and 4(b)] results in a droplet (indicating the presence of a continuous film at the bubble-solid interface even for small values of $h/R$) whilst bursting on a more hydrophilic film (PDMS and silicon dioxide) [Fig. 4(c) and 4(d)] results in an unstable liquid ring which stabilizes into a ring of droplets. The value of $h/R$ (for the wetting surfaces) for which the transition between forming a droplet after bursting and forming a ring of droplets after bursting was determined to be 0.245 (silicon dioxide) and 0.241 (PDMS). However, prior to bursting, it is important to note that the extrapolated values of $\theta_{bi}$ correspond very well to those values of $\theta_{b0}$ computed using Eq. (3) which does not require the presence of a continuous film.

Finally, Fig. 5 shows sessile bubbles of differing sizes wetting a liquid film composed of the same bubble solution. In general, the Plateau borders are larger – for a given bubble radius – than bubbles wetting a solid surface. The bubble contact angle is seen to diminish with bubble radius $r$. Fig. 6 shows a plot of $\theta_b$ versus bubble radius $r$. As $r$ is varied from 10 mm to 645 µm - $\theta_b$ changes from 68.4° to 33.6°. A plot of $\theta_b$ versus $h/R$ reveals itself to be linear [inset to Fig. 6] – as is case with bubbles wetting a solid surface - the intercept $\theta_{bi}$ is 91.9° and the slope $\alpha$ is -170.4. In terms of sessile bubbles wetting a liquid film of the same liquid [Fig. 5], the value of $\alpha$ predicted by the model[19] - assuming a perfectly hydrophilic surface - does not correspond well with the experimental data for sessile bubbles wetting a liquid surface. Note that, the experimental data for a hydrophilic surface given in Ref. [19] gives a slope $\alpha$ of ~90. Although the data here suggests a near-linear relationship – as is the case with bubbles wetting a solid surface [Fig. 3(b)] – the measured slope corresponds is 2.4 times the value of $\alpha$ predicted for wetting on a perfectly hydrophilic surface. In other words, the data indicates that wetting of a bubble on a perfectly hydrophilic solid surface is not the same as wetting of a bubble on a liquid film of the same liquid.

The author thanks Frank Hein (Pustefix) for discussions.

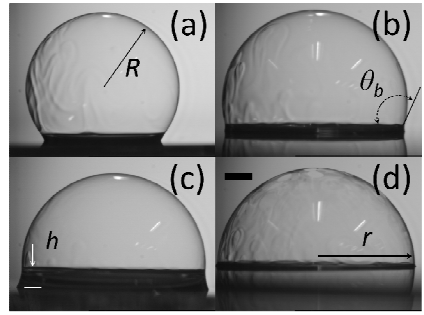

FIG. 2. Sessile bubbles wetting solid surfaces. (a) Fluorocarbon coated "black silicon", (b) Teflon® AF coated silicon wafer, (c) PDMS and (d) silicon dioxide. Scale bar = 1000 µm.

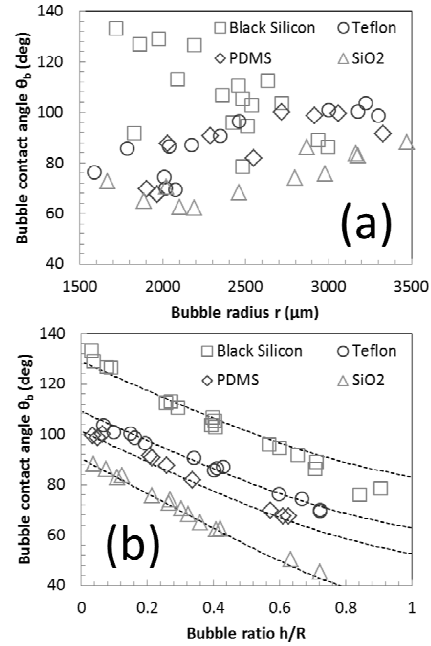

FIG. 3. Bubble contact angle $\theta_b$ versus the $h/R$ ratio for the surfaces tested. Fluorocarbon coated "black silicon" (open squares), Teflon® AF (open circles), PDMS (open diamonds) and silicon dioxide (open triangles). The dashed lines are analytical solutions.

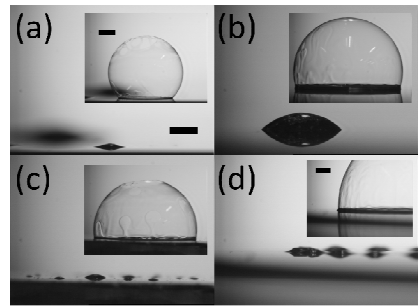

FIG. 4. Sessile bubbles bursting on solid surfaces. (a) Fluorocarbon coated "black silicon", (b) Teflon® AF coated silicon wafer, (c) PDMS and (d) silicon dioxide. Insets show outcome of bursting. Scale bars = 1000 µm.

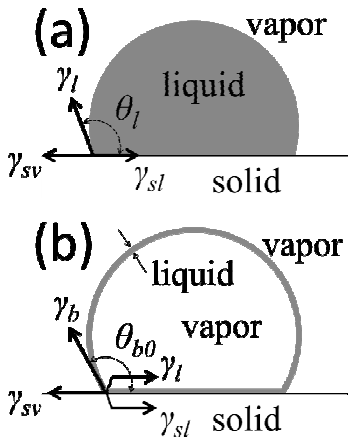

FIG. 1. Schematic diagrams of (a) a sessile droplet and (b) a sessile bubble resting on a solid surface.

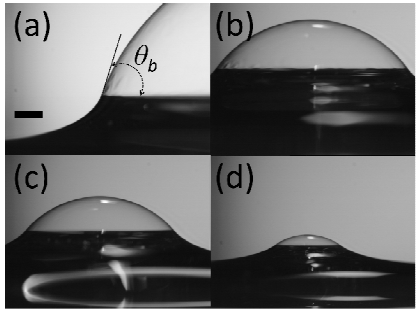

FIG. 5. Sessile bubbles wetting a liquid film. Scale bar = 1000 μm.

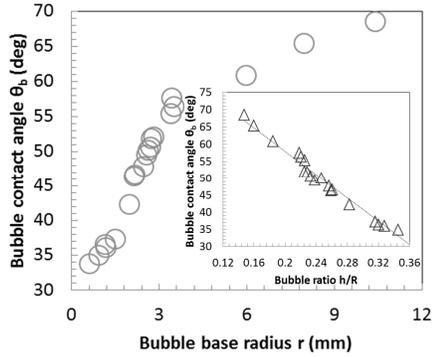

FIG. 6. Bubble contact angle $\theta_b$ versus $r$. Inset shows bubble contact angle $\theta_b$ versus the $h/R$ ratio.

|  | θl (exp.) | θbi (exp.) | α (exp.) | θb0 (calc.) | α (calc.) |
|---|---|---|---|---|---|
| Black Si | 111.9 (1.2) | 132.3 | -66.4 | 133.4 (0.8) | -56.3 |
| Teflon | 62.1 (2) | 106.6 | -50.7 | 105.4 (1.1) | -56.8 |
| PDMS | 52.8 (3) | 102.4 | -56.7 | 101.4 (1.2) | -59.5 |
| SiO2 | 9.6 (0.6) | 90.2 | -64.1 | 90.4 (0.1) | -68.9 |
| Liquid | 0 | 91.9 | -170.4 | 90 | -69.6 |

Table. Experimental and calculated values obtained from the study.